\documentclass{elsart}
\usepackage{amsmath}

\input psfig.sty
\input{tcilatex}

\begin{document}

\begin{frontmatter}

\title{On relativistic approaches to the pion self-energy in nuclear matter}

\author{L. B. Leinson$^{1}$ and A. P\'{e}rez$^{2}$}

\address{$^{1}$Departamento de F\'{i}sica Te\'{o}rica, Universidad de Valencia\\
46100 Burjassot (Valencia), Spain\\
and\\ Institute of Terrestrial Magnetism, Ionosphere and
Radio Wave Propagation
RAS, 142190 Troitsk, Moscow Region, Russia\\
$^{2}$Departamento de F\'{i}sica Te\'{o}rica and IFIC, Universidad de Valencia 
46100 Burjassot (Valencia), Spain \\
E-mail:\\ 
leinson@izmiran.rssi.ru\\
Armando.Perez@uv.es}

\begin{abstract}
We argue that, in contrast to the non-relativistic approach, a relativistic evaluation of the 
nucleon--hole and delta-isobar--nucleon hole contributions to the pion self-energy 
incorporates the s-wave scattering, which requires a more accurate evaluation. 
Therefore relativistic approach containing only these diagrams does not describe 
appropriately the pion self-energy in isospin symmetric nuclear matter. We conclude that, 
a correct relativistic approach to the pion self-energy should involve a more 
sophisticated calculation in order to satisfy the known experimental results on the 
near-threshold behaviour of the $\pi$-nucleon (forward) scattering amplitude. 

PACS number(s): 24.10.Cn; 13.75Cs; 21.65.+f; 25.70.-z 
\end{abstract}

\end{frontmatter}\newpage Originated from experiments with relativistic
heavy-ion collisions, considerable efforts from many theoretical groups were
made in relativistic approaches to the pion self-energy in isospin symmetric
nuclear matter (see e.g. \cite{Mao}, \cite{Herb}, \cite{Dmitriev}, \cite{Xia}%
, \cite{Lutz}, and references therein). Basically such calculations,
involving relativistic kinematics, are restricted to the contributions from
nucleon particle--hole ($ph$) and $\Delta $-isobar--nucleon hole ($\Delta $%
h) excitations in the medium, as given by the following diagrams:

\psfig{file=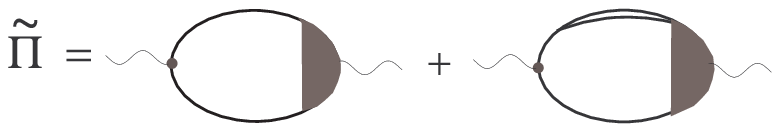} Here the $\Delta $-isobar is shown by double line.
The shadowed effective vertices for the pion interaction with nucleons and
deltas take into account the correlations in the medium. These vertices are
irreducible with respect to pion lines.

As has been pointed by many authors, such calculations yield a pion
self-energy $\tilde{\Pi}\left( \omega ,k\right) $, which, in the low-density
limit, does not reproduce exactly the pion self-energy obtained from the
non-relativistic reduction of the pion-nucleon and pion-delta Lagrangian.
The purpose of this Letter is to illuminate the reason of this discrepancy,
by showing that a relativistic approach containing only the above diagrams
does not describe the pion self-energy appropriately.

In the following we employ the widely used pseudovector interaction of pions
with nucleons and deltas. The corresponding Lagrangian density can be
written in the following form%
\begin{equation}
\mathcal{L}_{int}=\frac{f}{m_{\pi }}\bar{\psi}_{N}\gamma ^{\mu }\gamma _{5}%
\mathbf{\tau }\psi _{N}\partial _{\mu }\mathbf{\varphi }+\frac{f_{\Delta }}{%
m_{\pi }}\bar{\psi}_{\Delta }^{\mu }\mathbf{T}^{+}\psi _{N}\partial _{\mu }%
\mathbf{\varphi }+\frac{f_{\Delta }}{m_{\pi }}\bar{\psi}_{N}\mathbf{T}\psi
_{\Delta }^{\mu }\partial _{\mu }\mathbf{\varphi }.  \label{lp}
\end{equation}%
Here $\mathbf{\varphi }$ is the pseudoscalar isovector pion field, $m_{\pi }$
is the bare pion mass, and $f=0.988$ is the pion-nucleon coupling constant.
The excitation of the $\Delta $ in pion-nucleon scattering is described by
the last two terms in the Lagrangian with the Chew-Low value of the coupling
constant, $f_{\pi N\Delta }=2f$. The nucleon field is denoted as $\psi _{N}$%
, and $\psi _{\Delta }$ stands for the Rarita-Schwinger spinor of the $%
\Delta $-baryon. Here and below, we denote as $\mathbf{\tau }$ the isospin $%
1/2$ operators, which act on the isobaric doublet $\psi $ of the nucleon
field. The $\Delta $-barion is an isospin $3/2$ particle represented by a
quartet of four states. $\mathbf{T}$ are the isospin transition operators
between the isospin $1/2$ and $3/2$ fields.

The non-relativistic reduction of the pion-nucleon and pion-delta coupling,
given by Eq. (\ref{lp}), leads to an effective interaction Hamiltonian of
the form (see e. g. \cite{Mig71}, \cite{Eric}): 
\begin{equation}
\mathcal{H}_{int}=\frac{f}{m_{\pi }}\left( \mathbf{\sigma }\cdot \mathbf{%
\nabla }\right) \left( \mathbf{\tau }\cdot \mathbf{\varphi }\right) +\frac{%
f_{\Delta }}{m_{\pi }}\left( \mathbf{S}^{+}\cdot \mathbf{\nabla }\right)
\left( \mathbf{T}^{+}\cdot \mathbf{\varphi }\right) +h.c.,  \label{H}
\end{equation}%
where $\mathbf{\sigma }$ are the Pauli matrices, and $\mathbf{S}^{+}$ are
the transition spin operators connecting spin $1/2$ and $3/2$ states.

To show explicitly the above-mentioned problem we perform a relativistic and
non-relativistic calculation of the pion self-energy in a simple model,
where the $NN$, $N\Delta $, and $\Delta \Delta $ correlations are simulated
by phenomenological contact interactions with three Landau-Migdal
parameters, $g_{NN}^{\prime }$, $g_{N\Delta }^{\prime }$, $g_{\Delta \Delta
}^{\prime }$. (For details of the calculation see \cite{LP}.) Modern
experiments and theoretical estimates \cite{Wakasa}, \cite{Suz99}, \cite%
{Ari01} point out that $g_{N\Delta }^{\prime }$ must be essentially smaller
than $g_{NN}^{\prime }$ and $g_{\Delta \Delta }^{\prime }$. The most recent
analysis, reported in \cite{recent}, suggest $g_{NN}^{\prime }=0.6$, $%
g_{N\Delta }^{\prime }=0.24\pm 0.10$, $g_{\Delta \Delta }^{\prime }=0.6$.
While we will do not discuss possible deviations from this set of
Landau-Migdal parameters, let us investigate the behaviour of the pion
self-energy in this case.

\vskip0.3cm

\psfig{file=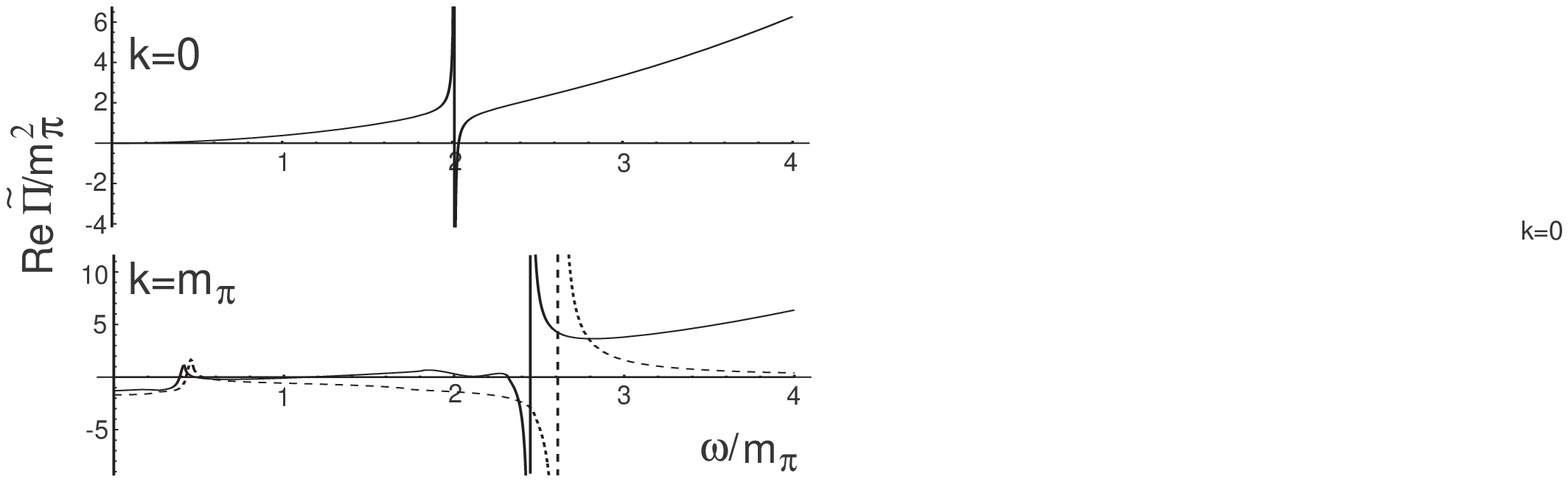} {Fig. 1. Pion self-energy in isosymmetric nuclear matter at 
saturation density $n=n_0$. The effective nucleon mass is taken to be $M^{\ast }=0.8M$. 
The solid line is obtained in the relativistic approach. The dashed line corresponds to
a standard non-relativistic calculation}

\vskip0.3cm In Fig. 1, the solid line presents the pion self-energy as
obtained in relativistic calculations. For a comparison, the dashed line
shows the pion self-energy calculated in the non-relativistic approach. As
one can see, even for $k=0$, we obtain a large discrepamcy. The relativistic
calculation results in an increasing of the pion self-energy along with $%
\omega $, while in the non-relativistic approach the pion self-energy
vanishes when $k=0$. One can easily find that the discrepancy arises even at
the lowest-order level:

\psfig{file=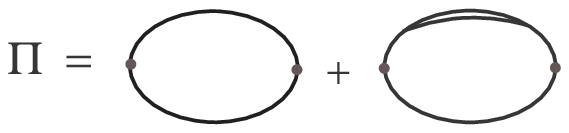}

Consider, for example, the particle-hole contribution, as given by the first
one-loop diagram. The relativistic evaluation yields (see e.g. \cite{Herb}):%
\begin{equation}
\func{Re}\Pi _{ph}\left( \omega ,k\right) =\frac{f^{2}}{\pi ^{2}}\frac{%
K_{\mu }K^{\mu }M^{\ast 2}}{m_{\pi }^{2}k}\int_{0}^{p_{F}}\frac{dpp}{%
\varepsilon }\ln \left| \frac{\left( K_{\mu }K^{\mu }-2kp\right)
^{2}-4\omega ^{2}\varepsilon ^{2}}{\left( K_{\mu }K^{\mu }+2kp\right)
^{2}-4\omega ^{2}\varepsilon ^{2}}\right| ,  \label{SeNN}
\end{equation}%
where $M^{\ast }$ is the effective nucleon mass, $\varepsilon ^{2}=M^{\ast
2}+p^{2}$, $p_{F}$ is the nucleon Fermi momentum, and $K^{\mu }=\left(
\omega ,\mathbf{k}\right) $ is the pion four-momentum.

It is instructive to analyse the low-density limit of this expression in
order to compare with the known non-relativistic form. At low density of
nucleons, $p_{F}/M^{\ast }\ll 1$, one has $\varepsilon \left( p\right)
\simeq M^{\ast }$. With this replacement, the integration can be performed
to give 
\begin{equation}
\func{Re}\Pi _{ph}\left( \omega ,k\right) =\frac{4f^{2}}{m_{\pi }^{2}}%
\,\left( \omega ^{2}-k^{2}\right) \left( \Phi _{0}\left( \omega
,k;p_{F}\right) +\Phi _{0}\left( -\omega ,k;p_{F}\right) \right) ,
\label{rph}
\end{equation}%
where 
\begin{equation}
\Phi _{0}\left( \omega ,k;p_{F}\right) =\frac{1}{4\pi ^{2}}\frac{M^{\ast 3}}{%
k^{3}}\left( \frac{1}{2}\left( a^{2}-k^{2}V_{F}^{2}\right) \ln \frac{a+kV_{F}%
}{a-kV_{F}}-akV_{F}\right)   \label{MigF}
\end{equation}%
is the Migdal function, with 
\begin{equation}
a=\omega +\frac{\omega ^{2}-k^{2}}{2M^{\ast }},\ \ \ \ \ V_{F}=p_{F}/M^{\ast
}.  \label{av}
\end{equation}%
This non-relativistic limit for the lowest-order particle-hole self-energy
has been obtained from relativistic kinematics. As given by Eq. (\ref{rph}),
for $\omega \ll 2M^{\ast }$ and in the limiting case of $k\rightarrow 0$, we
have: 
\begin{eqnarray}
\func{Re}\Pi _{ph}\left( \omega ,k\rightarrow 0\right)  &=&\frac{f^{2}}{%
M^{\ast }m_{\pi }^{2}}\,n\omega ^{2}  \notag \\
&&-\frac{f^{2}}{m_{\pi }^{2}}\,nk^{2}\left( \frac{1}{k^{2}/2M^{\ast }-\omega 
}+\frac{1}{k^{2}/2M^{\ast }+\omega }\right)   \label{NNnr}
\end{eqnarray}%
with 
\begin{equation*}
n=\frac{2p_{F}^{3}}{3\pi ^{2}}
\end{equation*}%
being the number density of isosymmetric nuclear matter. The corresponding
relativistic calculation of the lowest-order $\Delta h$ loop gives an
expression, which, in the low-density limit and $\omega \ll 2M^{\ast }$,
takes the following form \cite{LP}:%
\begin{equation}
\func{Re}\Pi _{\Delta h}\left( \omega ,k\rightarrow 0\right) =\frac{%
8f_{\Delta }^{2}}{9m_{\pi }^{2}}\,\frac{M^{\ast }+M_{\Delta }^{\ast }}{%
M_{\Delta }^{\ast 2}}n\omega ^{2}+\frac{8f_{\Delta }^{2}}{9m_{\pi }^{2}}\,%
\frac{M_{\Delta }^{\ast }-M^{\ast }}{\omega ^{2}-\left( M_{\Delta }^{\ast
}-M^{\ast }\right) ^{2}}nk^{2}  \label{NDnr}
\end{equation}%
A comparison of Eq. (\ref{NNnr}) and Eq. (\ref{NDnr}) with the
non-relativistic form of the lowest-order pion self-energy \cite{Eric}%
\begin{eqnarray}
\func{Re}\Pi ^{\mathrm{nr}}\left( \omega ,k\rightarrow 0\right)  &=&-\frac{%
f^{2}}{m_{\pi }^{2}}\,nk^{2}\left( \frac{1}{k^{2}/2M^{\ast }-\omega }+\frac{1%
}{k^{2}/2M^{\ast }+\omega }\right)   \notag \\
&&+\frac{8f_{\Delta }^{2}}{9m_{\pi }^{2}}\,\frac{M_{\Delta }^{\ast }-M^{\ast
}}{\omega ^{2}-\left( M_{\Delta }^{\ast }-M^{\ast }\right) ^{2}}nk^{2}
\label{Peric}
\end{eqnarray}%
shows that the relativistic evaluation results in additional contributions,
which do not vanish when $k\rightarrow 0$. In fact, one finds%
\begin{eqnarray}
&&\func{Re}\Pi _{ph}\left( \omega ,k\rightarrow 0\right) +\func{Re}\Pi
_{\Delta h}\left( \omega ,k\rightarrow 0\right) -\func{Re}\Pi ^{\mathrm{nr}%
}\left( \omega ,k\rightarrow 0\right)   \notag \\
&=&\left( \frac{f^{2}}{M^{\ast }m_{\pi }^{2}}\,+\frac{8f_{\Delta }^{2}}{%
9m_{\pi }^{2}}\,\frac{M^{\ast }+M_{\Delta }^{\ast }}{M_{\Delta }^{\ast 2}}%
\right) n\omega ^{2}  \label{dif}
\end{eqnarray}

To explain the origin of these terms we recall that the pion self-energy
represents the forward scattering amplitude of the pion in the medium. In
the non-relativistic theory, the above $ph$ and $\Delta h$ diagrams, taking
also into account the correlations in the medium, are known to reproduce
well the (forward) p-scattering amplitude in the isospin symmetric nuclear
matter, while the s-scattering contribution is known to be small. Due to the
non-relativistic reduction of $\pi NN$ interaction, as given by Eq. (\ref{H}%
), the s-wave scattering gives no contribution to the $ph$ and $\Delta h$
diagrams.

However, the relativistic form of the pion-nucleon and pion-delta
interactions, as given by Eq. (\ref{lp}), causes an s-wave contribution to
the above diagrams. Consider, for example, the $\pi NN$ interaction. At the
pion four-momentum $K^{\mu }=\left( \omega ,\mathbf{k}=0\right) $, these
couplings involve only the time component of the currents. In the
low-density limit, $p_{F}/M^{\ast }\ll 1$, the matrix element $\left\langle
N\left( p^{\prime }\right) \right| \bar{\psi}_{N}\gamma ^{0}\gamma _{5}%
\mathbf{\tau }\psi _{N}\left| N\left( p\right) \right\rangle $ is
proportional to $\mathbf{\sigma \cdot }\left( \mathbf{p+p}^{\prime }\right)
/2M^{\ast }$, and the non-relativistic reduction, Eq. (\ref{H}), of the $\pi
NN$ interaction implyes that the part of the scattering amplitude generated
by $\mathcal{L}_{int}$ at the second order vanishes for nucleons at rest.
However, this is not the case, if the time component of the interaction is
relativistically incorporated to the calculation of the $ph$ and $\Delta h$
diagrams. Integration over the nucleon momentum results in the contribution,
proportional to $\omega ^{2}$ as given by Eq. (\ref{dif}).

Thus, in contrast to non-relativistic approach, relativistic evaluation of
the $ph$ and $\Delta h$ contributions to the pion self-energy incorporates
the s-wave scattering. This means that a covariant relativistic evaluation
of the pion self-energy can not be restricted only to the $ph$ and $\Delta h$
diagrams. A correct calculation of the (forward) s-wave amplitude actually
requires the inclusion of many more complicated diagrams, since the
s-scattering is caused mostly by the short-distance interactions, $r_{0}\sim
M^{-1}$. When these diagrams are included, one can expect a very strong
cancellation of the s-wave contribution to the pion self-energy. Indeed, for
example, at the threshold $\omega =m_{\pi }$, from Eq. (\ref{NNnr}) and Eq. (%
\ref{NDnr}) we obtain the forward s-scattering amplitude as 
\begin{equation}
\frac{1}{n}\func{Re}\Pi \left( m_{\pi },0\right) =\left( \frac{f^{2}}{%
M^{\ast }}\,+\frac{8f_{\Delta }^{2}}{9}\,\frac{M^{\ast }+M_{\Delta }^{\ast }%
}{M_{\Delta }^{\ast 2}}\right) \simeq 1.3\;fm,  \label{srel}
\end{equation}%
The correlation effects, as well as a reasonable variation of the effective
nucleon (and delta) mass, do not change the order of magnitude of this
result. When the short-range correlations are taken into account we obtain 
\begin{equation}
\frac{1}{n}\func{Re}\tilde{\Pi}\left( m_{\pi },0\right) \simeq 1.11\;fm.
\label{srel1}
\end{equation}%
Chiral symmetry, however, imposes strong constraints on the near-threshold
behaviour of this isospin even amplitude \cite{Tom} and it is known
experimentally \cite{Sig}, \cite{Koch} to be much smaller, as compared to
that given by Eqs. (\ref{srel}) and Eq. (\ref{srel1}).

Thus one can conclude that, due to the s-wave contribution, a correct
relativistic approach to the pion self-energy would involve a more
sophisticated calculation, including some extra diagrams. A strong
cancellation of the s-wave contribution should be expected in this way, so
as to satisfy the above experimental results.

\section*{Acknowledgements}

This work was carried out within the framework of the program of Presidium
RAS '' Non-stationary phenomena in astronomy'' and was partially supported
by Spanish grants FPA 2002-00612 and AYA 2001-3490-C02.

\end{document}